\documentclass[aps,prl,twocolumn,groupedaddress,amsmath,amssymb,amsfonts,graphics]{revtex4}

\bibdata{Kondo}

\DeclareMathOperator{\Tr}{Tr}
\DeclareMathOperator{\T}{T}

\DeclareMathOperator{\diag}{diag}

\newcommand{\bracket}[1]{\langle #1 \rangle}

\usepackage{graphicx}
\usepackage{bm}
\usepackage{bbold}
\renewcommand{\vec}[1]{\bm{#1}}

\begin{document}

\title{Field-induced Antiferromagnetism and the Kondo Insulator-to-Metal Transition}

\author{K.~S.~D.~Beach}
\email[]{ksdb@mit.edu}
\homepage[]{web.mit.edu/ksdb/www}
\affiliation{Department of Physics, Massachusetts Institute of Technology, Cambridge MA 02139}

\author{Patrick A. Lee}
\affiliation{Department of Physics, Massachusetts Institute of Technology, Cambridge MA 02139}

\author{P. Monthoux}
\affiliation{Cavendish Laboratory, University of Cambridge, Cambridge CB3 0HE, United Kingdom}

\date{June 12, 2003}

\begin{abstract}
The Kondo lattice model, augmented by a Zeeman term, serves as a useful model of a Kondo insulator in an applied magnetic field. A variational mean field analysis of this system on a square lattice, backed up by quantum Monte Carlo calculations, reveals an interesting separation of magnetic field scales. For Zeeman energy comparable to the Kondo energy, the spin gap closes and the system develops transverse staggered magnetic order. The charge gap, however, survives up to a higher hybridization energy scale, at which point the canted antiferromagnetism is suppressed and the system becomes metallic. Quantum Monte Carlo simulations support this mean field scenario. An interesting rearrangement of spectral weight with magnetic field is found.
\end{abstract}

\maketitle

The unusual properties of heavy fermion materials are the result of strong interactions between their mobile $spd$- and tightly bound $f$-electrons. In the standard picture~\cite{Coleman}, a broad conduction band is intersected by a nearly flat band of renormalized core levels. Hybridization produces a band structure with extremely shallow dispersion near the band edge. When the chemical potential lies below the hybridization gap, the system is a metal characterized by a very large effective electron mass. When the chemical potential lies inside the gap, the system is an insulator. Since RKKY-mediated antiferromagnetism competes with this hybridization mechanism, heavy fermion systems typically live at the edge of magnetic instability~\cite{Doniach}.

The Kondo lattice model (KLM) is thought to provide an approximate description of the heavy fermion materials. In particular, at half-filling, local singlet formation dominates whenever the Kondo coupling $J$ exceeds some critical value $J_{\text{c}}$; RKKY antiferromagnetism wins out otherwise. The large $J$ regime is believed to describe such Kondo insulators as Ce$_3$Bi$_4$Pt$_3$, CeRhAs, CeRhSb, YbB$_{12}$, and SmB$_6$.

An applied magnetic field will interfere with the singlet--RKKY competition and, if strong enough, tilt the balance in favor of the antiferromagnetism. The Zeeman term, a field $B$ coupled to the total magnetic moment, favors triplet rather than singlet formation at each site. Accordingly, it suppresses the singlet amplitude and thus has the potential to stabilize an antiferromagnetic phase, even in the $J \!>\! J_{\text{c}}$ region. Moreover, the Zeeman splitting lifts the degeneracy of the spin up and spin down bands, shifting them with respect to one another and potentially closing the charge gap. Indeed, transport measurements of Ce$_3$Bi$_4$Pt$_3$ in high magnetic fields ~\cite{Jaime-short,Boebinger-short} indicate that its resistivity plummets to a low impurity-dominated value at a critical field on the order of 50 T.

In this Letter, we show by variational mean field and quantum Monte Carlo (QMC) calculations that as the applied field is ramped up the ground state of the KLM with Zeeman splitting (KLM+Z) transforms from Kondo insulator to conventional metal by way of a canted antiferromagnetic phase.

The KLM+Z describes a half-filled conduction band interacting with a lattice of localized spins $\{\hat{\vec{S}}_i\}$. The net magnetic moment at each site is coupled to an external field. The Hamiltonian is
$\hat{H} \!=\! \hat{H}_t \!+\! \hat{H}_{JB}$, where
\begin{equation}
\begin{split}
\hat{H}_t &= -t\sum_{\bracket{ij}}\bigl( c_i^\dagger c_j + c_j^\dagger c_i \bigr) ,\\
\hat{H}_{JB} &= J\sum_i \tfrac{1}{2}c^\dagger_i\vec{\sigma}c_i \cdot \hat{\vec{S}}_i 
-B\sum_i \Bigl(\tfrac{1}{2}c_i^{\dagger}\sigma^3c_i 
+ \hat{S}^3_i \Bigr) .
\end{split}
\end{equation}
Here, $t$, $J$, and $B$ are the hopping, Kondo coupling, and magnetic field, respectively. $c^{\dagger}$ ($c$) is the creation (annihilation) operator for the conduction electrons. The lattice is taken to be square---its bipartite nature being essential to our QMC scheme.

\emph{Variational Mean Field}---Take $\hat{\vec{S}} \!=\! \frac{1}{2}f^{\dagger}\!\vec{\sigma}\!f$, subject to the constraint $f^{\dagger}\!f \!=\! 1$, where $f^{\dagger}$ ($f$) is the creation (annihilation) operator of a fictitious, dispersionless $f$-band. At the mean field level, it is sufficient to enforce the constraint on average and require only that $\bracket{f^{\dagger}\!f} \!=\! 1$. In the $f$ language, the local spin singlet operator is $f^\dagger c$.

\begin{figure}
\includegraphics{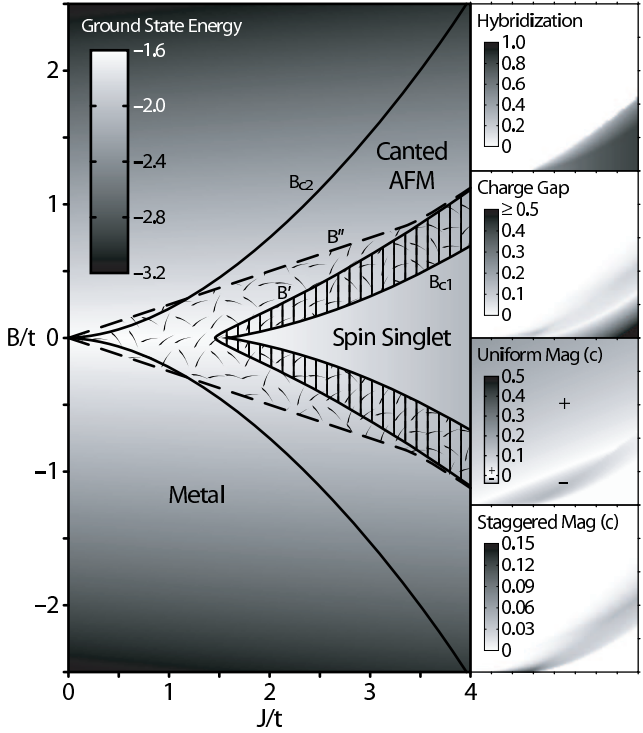}
\caption{\label{FIG:varmf} $T\!=\!0$ mean field, square lattice. (Left) The singlet phase has hybridization order only. Antiferromagnetic and hybridization order coexist in the striped region. The stippling indicates that the electron moments are directed opposite to the $B$ field. In the metallic phase, the local moments are saturated. (Right) $\bracket{f^\dagger c}$, $\Delta_{\text{c}}$, $\bracket{\tfrac{1}{2}c^\dagger_i\sigma^3c_i}$, and $\bracket{(-1)^i\tfrac{1}{2}c^\dagger_i\sigma^1c_i}$ in the upper half of the phase diagram.
}
\end{figure}

A comprehensive variational calculation of the ground state must incorporate all of the following: hybridization between the $c$ and $f$ bands (singlet formation), antiferromagnetism transverse to the field, and magnetism parallel to the field. A suitable trial wavefunction is the ground state of the mean field Hamiltonian
\begin{equation}
\begin{split}
\hat{H}_{\text{MF}} &= \hat{H}_t - \sum_i \Bigl( V^* f^{\dagger}_i c_i 
                                      + V c^{\dagger}_i f_i\Bigr)\\
                                      &\quad-\frac{M_f}{2} \sum_i (-1)^i c^{\dagger}_i \sigma^1 c_i
                                        -\frac{B_f}{2} \sum_i c^{\dagger}_i \sigma^3 c_i\\
                                      &\quad
                                      -\frac{M_c}{2} \sum_i (-1)^{i+1} f^{\dagger}_i \sigma^1 f_i -\frac{B_c}{2} \sum_i f^{\dagger}_i \sigma^3 f_i.
\end{split}
\end{equation}
It contains five symmetry-breaking terms, controlled by the variational parameters $\{V_m\} \!=\! \{V, M_c, M_f, B_c,  B_f\}$. The variational ground state energy $\mathcal{U}[\{V_m\}]$ is the expectation value of the exact Hamiltonian evaluated in the ground state of $\hat{H}_{\text{MF}}$. $\{V_m\}$ is then chosen such that $\mathcal{U}$ is minimized. Figure~\ref{FIG:varmf} depicts the resulting phase diagram.

\begin{figure}
\includegraphics{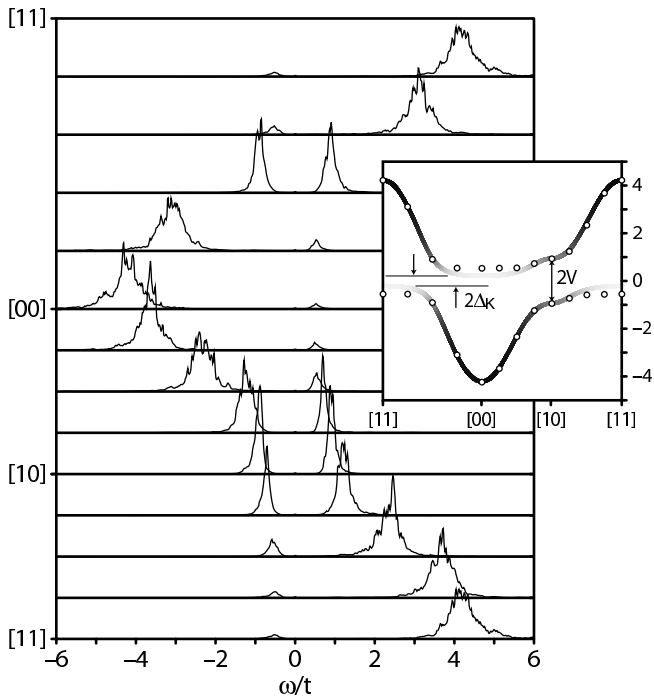}
\caption{\label{FIG:disp}Analytically continued data from the one-particle electron Green's function for $J/t \!=\! 1.7$, $\beta t \!=\! 14$ on an $8\!\times\!8$ lattice. Spectral functions $A(\vec{k},\omega)$ are plotted for wavevectors along high-symmetry lines in the Brillouin zone. (Inset) the peaks locations (open circles) are superimposed on the mean field band structure [Eq.~\eqref{EQ:banddisp}] with $V$ chosen to fit the band splitting at $\vec{k} \!=\! (\pi,0)$. The greyscale indicates the spectral weight (white--black $\leftrightarrow$ 0--1).}
\end{figure}

The singlet regime is characterized by nonzero hybridization ($V \!\neq\! 0$) and the absence of magnetic order ($M_{c,f} \!=\! 0$, $B_{c,f} \!=\! B$).  The energy levels, given by
\begin{equation} \label{EQ:banddisp}
E_{\vec{k}s}^n = \frac{1}{2}\Bigl( \varepsilon_{\vec{k}} -sB + n\sqrt{\varepsilon_{\vec{k}}^2 + 4V^2} \Bigr)
\end{equation}
with $\varepsilon_{\vec{k}} = -2t\sum_{l=1}^d\cos k_l$ (for a $d$-hypercubic lattice), are parameterized by the band index $n \!=\! \pm$ and spin projection $s \!=\! \pm \ (\uparrow,\downarrow)$. In the $B\!=\!0$ case (inset Fig.~\ref{FIG:disp}), the band separation takes its minimum, $2V$, on the surface $|k_1|\!+\cdots+\!|k_d| \!=\! \pi$ ($\varepsilon_{\vec{k}} \!=\! 0$). The band gap, however, is indirect: promoting a quasiparticle from the top of the lower band [at $\vec{Q}\!=\!(\pi,\ldots,\pi)$] to an arbitrary momentum state in the upper band costs $\omega^{\text{qp}}(\vec{k}) \!=\! E^+_{\vec{k}}\!-\!E^-_{\vec{Q}}$; around its minimum, $\omega^{\text{qp}}(\vec{k}) \!\approx\! 2\Delta_{\text{K}} +(1/2m^*)|\vec{k}|^2$, where $4V^2 \!=\! 2\Delta_{\text{K}}\bigl(2\Delta_{\text{K}} \!+\! W \bigr)$ and $W \!=\! 4dt$ is the noninteracting bandwidth. The so-called Kondo energy, $\Delta_{\text{K}} (<V)$, sets the scale for both the charge gap ($\Delta_{\text{c}} \!=\! 2\Delta_{\text{K}}$) and the ground state energy shift ($\mathcal{U}[V] \!-\! \mathcal{U}[0] \!\approx\! -\Delta_{\text{K}}$).

Now consider $B\!\neq\!0$. Throughout the singlet phase, $V$ is independent of the applied field. Thus, according to Eq.~\eqref{EQ:banddisp}, the hybridized bands simply shift with respect to one another in response to the applied field and the charge gap decreases monotonically: $\Delta_{\text{c}} \!=\! 2\Delta_{\text{K}}\!-\!|B|$. Before the charge gap closes completely, however, magnetic order sets in ($|B| \!=\! B_{\text{c}1} \!\lesssim\! 2\Delta_{\text{K}})$. The localized spins develop  a uniform moment directed with the field and a staggered moment perpendicular to it. The electrons do the same but are ``effectively diamagnetic,'' canting \emph{against} the field. As $|B|$ increases, $V$ begins to falls off and vanishes at $|B| \!=\! B'$. When $|B| \!=\! B'' \!\approx\! \max(J/4,B')$, $\bracket{\tfrac{1}{2}c^\dagger_i\sigma^3c_i}$ changes sign. When $|B| \!=\! B_{\text{c}2}$, the local spins become saturated and the antiferromagnetism dies out; the conduction electrons are decoupled from the quenched local spins and propagate freely.

\begin{figure}
\includegraphics{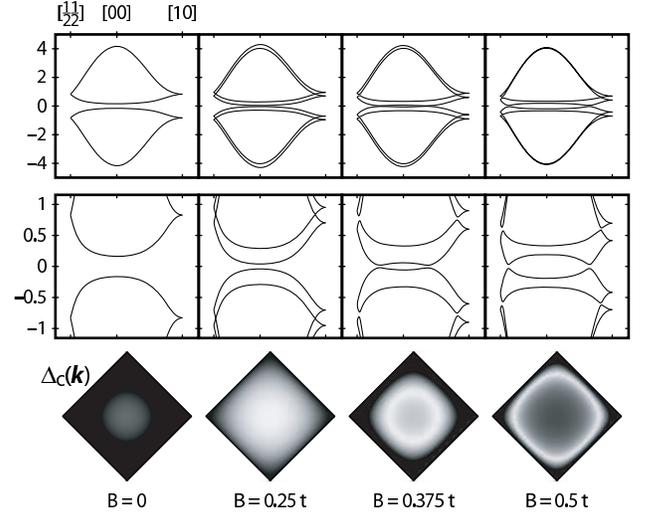}
\caption{\label{FIG:bands} $J/t \!= \!3$, $B_{\text{c}1}/t \!\doteq\! 0.323$. (Top row) Mean field band structure, folded into the reduced Brillouin zone, as a function of applied magnetic field. (Middle) Same, magnified to show the evolution of the band gap. (Bottom) The greyscale (white--black $\leftrightarrow$ $0$--$0.5t$) indicates the wavevector-dependent gap magnitude. For $|B|\!<\!B_{\text{c}1}$, $\Delta_{\text{c}}(\vec{k})$ is a minimum at $\vec{k}\!=\!0$, and for $|B|\!>\!B_{\text{c}1}$, at an expanding ring of points.}
\end{figure}

Naively, one might have expected the charge gap to close at $|B|\!=\!2\Delta_{\text{K}}$. 
As shown in Fig.~\ref{FIG:bands}, however, the incipient antiferromagnetism prevents the level crossing by mixing the $(c,f)_{\vec{k}\uparrow}$ and $(c,f)_{\vec{k}+\vec{Q}\downarrow}$ bands. The charge gap has the behavior shown in Fig.~\ref{FIG:varmf} (sideset, right): it decreases linearly with $|B|$ in the singlet phase, resurges in the canted antiferromagnetic phase, and finally vanishes in the metallic phase.

An important feature of the metallic transition is that the charge gap does not close at the center of the reduced (mod $\vec{Q}$) Brillouin zone. As the system is driven through the antiferromagnetic phase ($B_{\text{c}1} \!<\! |B| \!<\! B_{\text{c}2}$), the gap migrates from $\vec{k} \!=\! 0$ out to the zone edge before closing. A consequence is that the hybridization energy and not the Kondo energy determines the robustness of the insulating state. This leads to a separation of energy scales: the spin gap closes when $|B| \!\sim\! 2\Delta_{\text{K}}$, whereas the charge gap closes when $|B| \!\sim\! \sqrt{2\Delta_{\text{K}}W}$.

\emph{Quantum Monte Carlo}---The partition function of the system can be written as
\begin{equation} \label{EQ:partition}
Z = \int d\eta d\bar{\eta}\, e^{\bar{\eta}\eta}\langle \eta \rvert
         \prod_{k=1}^L \hat{P}\int \mathcal{D}\chi_k\, e^{-\epsilon \hat{H}[\chi_k]}\hat{P}
         \lvert -\eta \rangle ,
\end{equation}
where $k$ labels the slices of discretized imaginary time ($\epsilon \!=\! \beta/L$), the Grassman vector $\eta \!=\! \begin{pmatrix}c_{\uparrow} & c_{\downarrow} &
                                        f_{\uparrow}  & f_{\downarrow}\end{pmatrix}$
 holds the four fermionic species, and the operator $\hat{P}$ projects out states that violate the single occupancy requirement of the $f$ electrons. $\hat{H}[\chi_k]$ is bilinear and related to the Hamiltonian by a Hubbard-Stratonovich decomposition of the Kondo interaction in the hybridization channel.

Replacing $e^{-\epsilon \hat{H}[\chi_k]}$ by its coherent state representation puts a set of Grassman states at each time slice $k$. The matrix elements $\langle \eta_k \lvert \hat{P} \rvert \eta_{k+1} \rangle$ can be handled by introducing a U(1) gauge field $z_{k,k+1}$ living on the temporal links. Integrating out the Grassman fields gives 
$Z \!=\! \int \!\mathcal{D}\chi \mathcal{D}z \,e^{-S[\chi,z]}$, where $S \!=\! \sum_{i,k} |\chi_{i,k}|^2\!-\!\Tr \ln M[\chi,z]$ 
is the action of a lattice gauge theory 
in $d\!+\!1$ dimensions. By exploiting the particle-hole symmetry of the Hamiltonian and the bipartite nature of the lattice, 
it is possible to transform to a gauge in which $M$ is positive definite. Specifically, under $c_{i\uparrow}\!\rightarrow\! c_{i1}$, $c_{i\downarrow}\!\rightarrow\! (-1)^i\bar{c}_{i2}$, $f_{i\uparrow}\!\rightarrow\! f_{i1}$, $f_{i\downarrow}\!\rightarrow\! (-1)^{i\!+\!1}\bar{f}_{i2}$, the gauge theory acquires a block diagonal form, $M = \diag(M_1,M_2)$ with $M_2 = M_1^\dagger$, so that $\det M = |\det M_1|^2$. Since the total magnetic moment transforms as $\sum_i\bigl( \bar{c}_i\sigma^3c_i\!+\!  \bar{f}_i\sigma^3f_i \bigr)\!\rightarrow\!\sum_i\sum_{s=1,2}\bigl( \bar{c}_{is}c_{is}\!+\!\bar{f}_{is}f_{is} \bigr)$, the positivity of $M$ is preserved even for $B\!\neq\!0$.

A general $n$-particle correlation function of the form
$Z^{-1}\!\!\int\! \mathcal{D}\chi \mathcal{D}z \, M^{-1}_{i_1i'_1;\tau_1\tau'_1} \cdots M^{-1}_{i_ni'_n;\tau_n\tau'_n} e^{-S[\chi,z]}$ is evaluated via stochastic sampling~\cite{Metropolis} by interpreting $\mathcal{P}[\chi,z] \!=\! Z^{-1}e^{-S[\chi,z]}$ as a probability weight.  (Since $\det M \!>\! 0$, there is no fermion sign problem.) The sampling algorithm ensures that any particular configuration $\chi, z$ is visited with probability $\mathcal{P}[\chi,z]$. Thus, computing correlation functions amounts to binning and tabulating $M_{ii';\tau\tau'}^{-1}$ for a large series of independent field configurations. Updates are effected by a more sophisticated version of the algorithm introduced by Blankenbecler, Scalapino, and Sugar~\cite{Blankenbecler}. Implementation details and other technical issues are discussed at length elsewhere~\cite{ksdb-future}. Our method is related to that of Caponi and Assaad, who have carried out QMC for the KLM in zero field~\cite{Capponi}. The reliability of our code is verified by comparison with their results: \emph{e.g.}, our computed critical coupling, $J_{\text{c}}/t \!=\! 1.47 \!\pm\! 0.08$, is consistent with their value of $1.45 \!\pm\! 0.05$. 

The one-particle electron Green's function $G(\vec{k},\tau) \!=\! \bracket{\T [ c_{\vec{k}}(0)c_{\vec{k}}^\dagger(\tau)]}$ corresponds to $\int\! \mathcal{D}\chi \mathcal{D}z\, M_{\vec{k};0\tau}^{-1} \mathcal{P}[\chi,z]$.
Written in terms of its spectral function $A(\vec{k},\omega)$, the Green's function has the form of a linear functional
$G(\vec{k},\tau) \!=\! \int\! d\omega \, K(\tau,\omega)A(\vec{k},\omega) \!=\! \mathbf{K}[A(\vec{k},\omega)]$ with kernel
$K(\tau,\omega) \!=\! e^{-\omega\tau}/(e^{-\beta\omega}\!+\!1)$. To extract $A(\vec{k},\omega)$, we perform a (systematically regularized, see Ref.~\onlinecite{ksdb-future}) functional inversion $A(\vec{k},\omega) \!=\! \mathbf{K}^{-1}[G(\vec{k},\tau)]$. Figure~\ref{FIG:disp} shows the zero-field spectrum in the singlet phase ($J \!>\! J_{\text{c}}$). Note that the spectral peaks trace out two bands separated by a small gap at the chemical potential. The lower and upper bands exhibit (heavy fermion) regions of low spectral weight and nearly flat dispersion in the vicinity of $\vec{k} \!=\! (\pi,\pi)$ and $\vec{k} \!=\! 0$, respectively.

QMC results confirm that the KLM+Z system is well-described by its mean field theory. Several signature features are observed in the simulations: a robust singlet phase ($J \!>\! J_{\text{c}}$, $|B| \!<\! B_{\text{c}1}$), electronic moments directed against the field ($B_{\text{c}1} \!<\! |B| \!<\! B''$), transverse staggered magnetic order ($B_{\text{c}1} \!<\! |B| \!<\! B_{\text{c}2}$), and a high-field insulator-to-metal transition ($|B| \!=\! B_{\text{c}2}$) in which the charge gap closes near the reduced zone edge.

\begin{figure}
\includegraphics{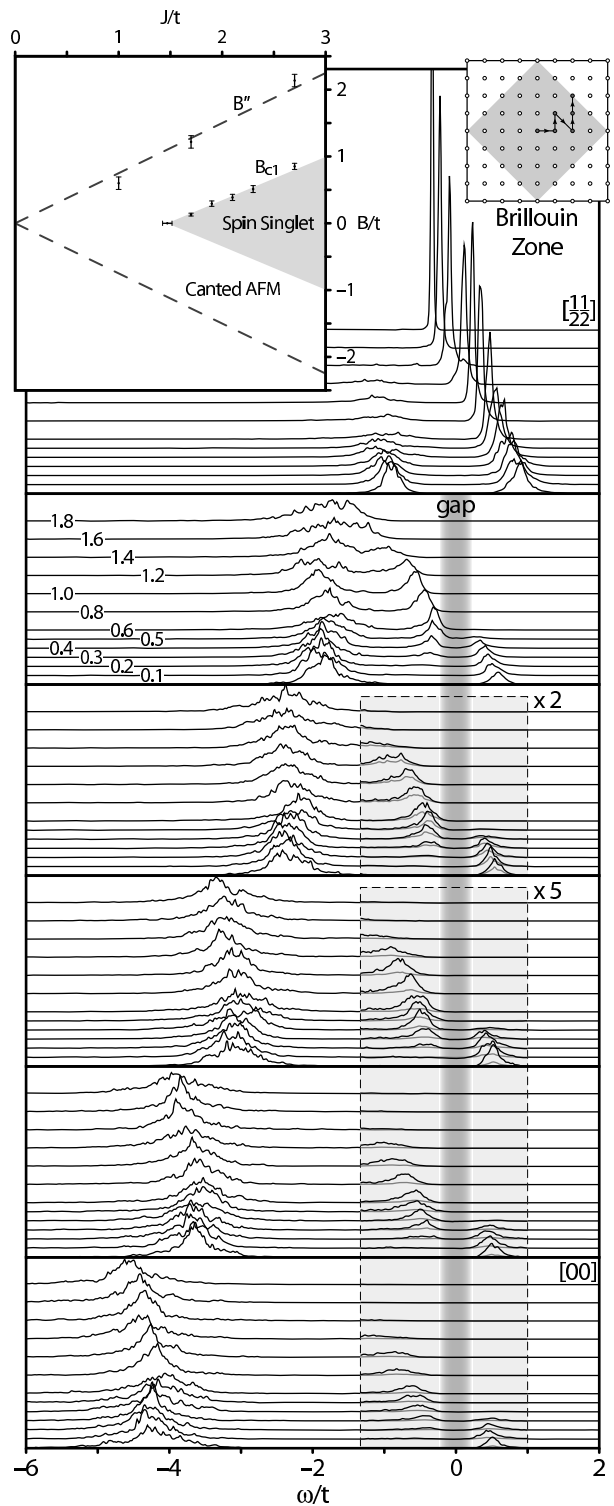}
\caption{\label{FIG:qmc} $J/t \!=\! 1.7$, $\beta t \!=\! 14$, $8\!\times\!8$ lattice. Spectral functions $A_{\uparrow}(\vec{k},\omega)$ are plotted over a range of field values ($0 \!\le\! B/t \!\le\! 1.8$ offset) for a series of wavevectors (identified in the top right inset). Smaller peaks have been scaled by the indicated magnification factor. Note that, instead of shifting smoothly through zero in a magnetic field, the spectral weight jumps across the gap, except for $\vec{k}$ on the reduced zone boundary. (Top left inset) A partial phase diagram. Electron moments are directed against the field between $B_{\text{c}1}$ and $B''$.}
\end{figure}

The inset in Fig.~\ref{FIG:qmc} depicts the phase boundary between the singlet and canted antiferromagnetic phases; it also marks where $\bracket{\tfrac{1}{2}c^\dagger_i \sigma^3 c_i}$ changes sign. The main plot shows in six panels the spectral function of a spin up electron evaluated at a series of wavevectors snaking through the Brillouin zone from $\vec{k} \!=\! 0$ to $\vec{k} \!=\! (\pi/2,\pi/2)$; the panels are arranged (from bottom to top) in order of increasing distance from the zone center.

As per Eq.~\eqref{EQ:banddisp}, the spectra exhibit a double peak structure and drift leftward (lowering energy) as $B$ increases from zero. For small $\vec{k}$, there is a primary peak near the non-interacting particle energy $\omega \!=\! \varepsilon_{\vec{k}}$ and a secondary peak near $\omega \!=\! \Delta_{\text{K}}$. The leftward drift of the secondary peak is interrupted by the growth of antiferromagnetic correlations that protect the gap. \emph{The spectral weight rearranges by hopping over the forbidden region}. There is no weight at the chemical potential, so the system remains insulating even when $|B| \!=\! 2\Delta_{\text{K}}$.

At $\vec{k} \!=\! (\pi/2,\pi/2)$, there are two equally weighted peaks at $\pm V$. As $B$ is ramped up, spectral weight from the lower peak is transferred to the upper, unoccupied peak (this accounts for $\bracket{\tfrac{1}{2}c^\dagger \sigma^3 c} \!<\! 0$), which grows increasingly sharper as it drifts leftward and crosses the chemical potential (fixed at $\mu = 0$ by particle-hole symmetry). On the approach to the metallic transition, the spectral weight of the lower peak is exhausted and $A(\vec{k},\omega)\!\rightarrow\! \delta\bigl(\omega\!-\!B/2)$, its free electron value; the antiferromagnetic correlations die out here as well. Contrary to mean field prediction, the closing of the gap does not coincide exactly with saturation of the local moments.

\emph{Summary}---We have investigated the effect of an applied magnetic field on the Kondo insulator ground state, using mean field calculations (in the thermodynamic limit) and QMC simulations (on small lattices) to characterize the phases of the KLM+Z. From these numerical results, a consistent picture emerges: (i) In the $J \!>\! J_{\text{c}}$ local singlet phase, the hybridized bands respond to the applied field by shifting with respect to one another and reducing the charge gap. (ii) For $|B| \!\sim\! 2\Delta_{\text{K}}$, the spin gap closes and the local singlet phase gives way to an antiferromagnetic phase in which the local spins cant \emph{with} and the electron moments cant \emph{against} the applied field. The hybridization-supported charge gap, which had nearly closed, reinflates, driven now by the magnetic order. (iii) At larger fields, there is a crossover at which the electron moments begin to cant with the field. (iv) For $|B| \!\sim\! \sqrt{2\Delta_{\text{K}}W}$, the antiferromagnetism vanishes and the system becomes metallic.

This work was partially supported by the National Computational Science Alliance under grant DMR020009. Computations were performed on the SGI Origin 2000 and IA-64 Linux Cluster. The authors thank T. M. Rice for helpful discussions.

\bibliography{Kondo}

\end{document}